\documentclass[pra,twocolumn]{revtex4}

\usepackage{amsmath, amsthm, amssymb}
\usepackage{graphicx}
\usepackage{dcolumn}
\usepackage{bm}

\newcommand{\de}[1]{\left( #1 \right)}

\renewcommand{\Re}[1]{{\mathrm{Re}}\de{#1}}

\newcommand{\ket}[1]{\left| #1 \right\rangle}

\newcommand{\tr}{\mathrm{Tr}}

\newcommand{\identity}{\openone}
\newcommand{\cala}{$\cal{A}$ }

\newcommand{\eg}{{\it{e.g.}}}
\newcommand{\ie}{{\it{i.e.}}}

\newcommand{\be}{\begin{equation}}
\newcommand{\ee}{\end{equation}}
\newcommand{\bea}{\begin{eqnarray}}
\newcommand{\eea}{\end{eqnarray}}

\begin{document}

\title{Area laws and entanglement distillability of thermal states}

\author{Daniel Cavalcanti$^1$, Alessandro Ferraro$^1$, Artur Garc\'ia-Saez$^1$
and Antonio Ac\'\i n$^{1,2}$}
\affiliation{$^1$ICFO-Institut de Ciencies Fotoniques,
Mediterranean
Technology Park, 08860 Castelldefels (Barcelona), Spain\\
$^2$ICREA-Instituci\'o Catalana de Recerca i Estudis Avan\c cats,
Lluis Companys 23, 08010 Barcelona, Spain}

\begin{abstract}
  We study the entanglement distillability properties of thermal
  states of many-body systems. Following the ideas presented in [D.
  Cavalcanti et al., arxiv:0705.3762], we first discuss the appearance
  of bound entanglement in those systems satisfying an entanglement
  area law. Then, we extend these results to other topologies, not
  necessarily satisfying an entanglement area law. We also study
  whether bound entanglement survives in the macroscopic limit of an
  infinite number of particles.
\end{abstract}


\maketitle

\section{Introduction}

Quantum Information Theory was born as a framework capable of
describing the physics behind the processing, storage, and exchange of
information under the rules of Quantum Mechanics \cite{nielsen}.
Since it uses a very general and rather abstract approach, we do not
need to talk about specific systems, interactions, or Hamiltonians,
but instead we can play with qubits, channels, logic gates, etc.
Beyond information purposes, this abstraction has also been proven
very useful when studying relevant questions in other subareas of
physics, such as Condensed Matter \cite{EntMBS}, Statistical Mechanics
\cite{PopShoWint}, Quantum Optics \cite{QO}, or Astrophysics
\cite{Ast}.

In a recent work \cite{nos}, we have discussed the entanglement
distillability properties of thermal states of some quantum many-body
models with local interactions. We have shown the existence of a
temperature range for which no pure-state entanglement can be
distilled from the system despite being entangled. This type of
irreversible quantum correlations is also known as bound entanglement
\cite{Hor}.  This result, which can be valid for systems of arbitrary
size, was connected to the so called \emph{entanglement-area laws}, a
typical feature of these systems that relates the entanglement of two
distinguished regions to the area between them \cite{area}. In the
present contribution, we extend these ideas by considering two ways of
addressing area laws and discussing the role they play in the
appearance of bound entanglement. We also show that bound entanglement
can appear in systems with different topologies, not necessarily
fulfilling an entanglement area law.

Let us start by saying few words about the main subjects considered
here: distillability and area laws. It is well known that some
entangled mixed states have the property of being distillable. This
means that if one considers many copies of such a state, it is
possible to purify the entanglement into a smaller number of entangled
pure states using local operations and classical communication (LOCC)
\cite{Benn1,Benn2}. Although all two-qubit and one-qubit-one-qutrit
states are distillable, there are states in higher dimension for which
no LOCC operation can purify entanglement \cite{Hor}. These states are
called bound entangled. Specifically, in the case of a multipartite
system an entangled state of $n$ parties is bound entangled whenever
the $n$ parties cannot distill any pure-state entanglement out of it by
LOCC.

The first criterion able to detect bound entanglement was given by the
Peres criterion \cite{Peres}: all distillable states have a
non-positive partial transposition. Thus, finding an entangled state
with positive partial transposition (PPT) guarantees
non-distillability. In this paper we will use a quantitative version
of this criterion, namely the negativity $E_N(\rho)$ \cite{VidWer}, to
analyze the distillability properties of thermal states. The
negativity is defined as follows:
\begin{equation}\label{negat}
E_N(\rho)=\sum_{\lambda_i<0}|\lambda_i|,
\end{equation}
being $\lambda_i$ the eigenvalues of $\rho^{T_A}$, the partial
transposition of $\rho$ with respect to a given part $A$ of the
system. In other words, $E_N(\rho)$ is given by the sum of the
absolute values of the negative eigenvalues of $\rho^{T_A}$.  So if
the negativity of a state is zero, its partial transposition is
positive, and the system is non-distillable (either separable or bound
entangled), i.e.,
$$E_N(\rho)=0~\Rightarrow~\rho~\text{is non-distillable,}$$
$$E_N(\rho)>0~\Rightarrow~\rho~\text{is entangled.}$$
Another useful and related quantity used throughout this paper is the
logarithmic negativity $E_l(\rho)$, given by
$E_l(\rho)=\log_2(1+E_N(\rho))$. Clearly the implications above also
apply to $E_l(\rho)$.

Consider a system in a pure state and a bipartition of it into two
complementary subsystems. It is well known that for bipartite pure
states the entropy of one of the reduced systems uniquely measures
its entanglement \cite{Ben96,PR97}. As in Thermodynamics, one may
expect that the entropy increases with the volume of the reduced
state, but curiously this is not the case for the ground state of
many models considered so far
\cite{EntMBS,Vidal,Latorre,Aud,PED+05,CEP07,areath}. In fact, it
is instead seen that the entanglement between the two
complementary regions scales at most as their boundary area (in
non-critical situations), a behavior that is generally named
entanglement-area law. The above mentioned works approach
entanglement-area laws in a variety of ways. Here, we
will in particular focus on two different approaches. On one hand,
we will address the entanglement-area relation by keeping fixed
the size of the system while changing the geometry of the
bipartition. For example, in the case in which the distinguished
subsystem is contiguous and the total system is considered in the
macroscopic limit such an approach is usually named as block
entropy. When a system obeys an area law under this approach we
will say that an area law of type I is fulfilled. On the other
hand, one can consider also the opposite approach. Namely, how the
entanglement scales when a given partition is kept fixed while
changing the size of the system. An example of such an approach is
given, again for contiguous partitions, in Ref.\cite{CEP07}, where
an area law in the half-half partition (i.e., when a contiguous
group composed by half of the particles belongs to the first
subsystem and the other half to the second one) is established.
In this case we will say that an area law of type II is obeyed.
We consider both alternatives here, showing in which sense they
are not equivalent and how they help in enlightening the
appearance of bound entanglement.

The paper is organized as follows. In Sec. II we will point out the
role that area laws play in the appearance of bound entanglement at
finite temperatures. Then, in Sec. III we will focus on harmonic
oscillator systems and analyze the emergence of different forms of
area laws by studying different partitions and system sizes. We also
consider a configuration where an area law is not seen in its simpler
form and discuss the existence of bound entanglement for this case. In
Sec. IV we extend our analysis to the emergence of bound entanglement
also for spin$-\frac{1}{2}$ systems, obtaining very similar results as
for harmonic systems. Sec. V is devoted to concluding remarks.

\section{Bound entanglement and area laws}

In Ref. \cite{nos} we considered systems exhibiting area laws and
suggested that they are good candidates for presenting thermal bound
entanglement in the macroscopic limit. This comes from the fact that
when we increase the systems' size the ground-state negativity for
some partitions increases (e.g. for the even-odd cut, where even
particles belong to one subsystem and odd particles to the
complementary subsystem - see \cite{Aud}) while for other type of
partitions it saturates (e.g.  in the half-half geometry). It is then
natural to expect that, when temperature is added to the system, the
negativity in the half-half partition vanishes for a lower temperature
than in the even-odd geometry. In other words, one expects that
different partitions of the system become PPT at different
temperatures (that we call threshold temperatures). An important
feature of the considered systems is that they are translationally
invariant and then all half-half partitions are equivalent. This
observation implies that when the negativities of the half-half
partitions are null no pure-state entanglement can be distilled by
LOCC \cite{nos}.  On the other hand, there is still a temperature
range where the negativity for the even-odd partition is strictly
positive, which is enough to prove bound entanglement.

As already mentioned, a lot of efforts have been devoted to the
study of entanglement-area laws for the ground state of various
physical systems. Much less is known about equivalent laws for the
case of finite temperature, where the state of the system is in a
thermal mixture. Partly, this is due to the complexity
characterizing the structure of entanglement for mixed states. For
example, one of the known exact results concerns again the entropy
of a contiguous subsystem \cite{ent_th}. However, for mixed
states, this quantity is no longer a measure of entanglement.
To the best of our
knowledge, the most general result at finite temperature has been
derived recently by Wolf et al. \cite{areath} and gives a bound
to the mutual information between two complementary subsystems. We
recall that the mutual information is a measure of the total
amount of correlations, both classical and quantum. Hence, it
trivially gives an upper bound to the entanglement. An important
feature of such a bound is that the dependence with the
temperature $T$ and the area \cala between the complementary
regions is factorized (in particular the bound scales linearly
with \cala and $T^{-1}$). On the basis of this result, one may
argue that a trivial area law, \ie, in a factorized form, holds for
the negativity $E_N$ whenever:
\begin{equation}
\label{sal}
E_N\leq f(T)g(\cal{A})\,,
\end{equation}
where $f(T)$ and $g(\cal{A})$ are generic functions also depending on
the parameters of the system. In particular, we assume that these
functions do not depend on the way of partitioning the system. As
said, this type of relation holds for the bound found in Ref.
\cite{areath}, in the case of the mutual information. In case the
inequality in the previous formula becomes an equality, we refer to
this form of area law as a {\em strict} area law.  Notice that such a
form of area law has been shown to hold for a nearest-neighbor
harmonic ring at finite temperature, in the case of even-odd partition
\cite{nos}. There, numerical evidences of the validity of a strict
area law have been reported also for half-half partitions as well as
for the analogous cases in a spin ring.

Let us now consider how a strict area law affects the existence of
bound entanglement. As said, the key ingredient in the recipe above in
order to show the presence of bound entanglement is that different
partitions of a systems become PPT at different temperatures. As a
consequence, if a strict area law of type I holds then no bound
entanglement should be expected, since all partitions become PPT
at the same temperature, namely when $f(T)=0$ \cite{noteppt}.
Notably, we have found
no system showing this behavior, even considering models that exhibit
the same entanglement for different partitions in the ground state, as
we will report in detail in the next Sections. On the contrary we
observed, for any system taken into account, that a strict area law of
type I does not hold and that different partitions become PPT at
different threshold temperatures (in any case, the bound given in
Ref.~\cite{areath} is of course not violated). This means that for a
fixed system size, there is a temperature range for which
bound entanglement is present.

Now, one may wonder whether this holds
also in the macroscopic limit of an infinite number of particles.
In this case, it is the validity of a
strict area law of type II that does give the key to a positive
answer.  In fact, it ensures that the threshold temperatures for each
of the partitions chosen to reveal the presence of bound entanglement
(\eg, even-odd and half-half partitions) stay constant as the size of
the system increases.  In particular, it ensures that the range of
temperatures for which bound entanglement is present survives up to a
macroscopic level. We will see explicitly that this is
actually the case for some models consisting of harmonic chains
and that the same behavior seems to be valid also for spin chains.
In order to show that different scenarios may arise in other
topologies, we also considered systems with a star configuration. There,
we will see that a strict area law of type II is no more
valid. The next sections are devoted to show explicitly the
ideas explained above.

\section{Harmonic Oscillators}

Consider a system consisting of $N$ harmonic oscillators interacting
via the following Hamiltonian:
\begin{equation}\label{Hosc}
H=\frac{1}{2}\left(\sum_i p_i^2+\sum_{i,j}x_i V_{i,j} x_j\right),
\end{equation}
where $x_i$ and $p_i$ represent position and momentum operators for
each oscillator respectively ($i=1,\dots,N$).

The matrix $V$ describes both the on-site interaction (given by the
diagonal elements) and the coupling between oscillators $i$ and $j$
(non-diagonal terms). This Hamiltonian is quadratic in the canonical
coordinates and the oscillators are coupled through their position
degrees of freedom which sets both the ground and the thermal states
to be Gaussian. In this way these states are completely determined
by their covariance matrix $\gamma$ defined as follows. Take the
vector $S=(x_1,\dots,x_N,p_1,\dots,p_N)$, we then have
\begin{equation}
\gamma_{kl}=\Re{\tr\{\varrho[S_k-\bar S_k][S_l-\bar S_l]\}}
\end{equation}
where $\varrho$ is the density matrix of the state and $\bar
S_k=\tr(\varrho S_k)$. If we consider the thermal state
$\varrho=\exp[-H/T]/\tr\{\exp[-H/T]\}$ at temperature $T$, the
corresponding covariance matrix is given by \cite{Aud}:
\begin{equation}
\gamma(T)=[V^{-1/2}W(T)]\oplus[V^{1/2}W(T)]\,,
\end{equation}
where
\begin{equation}
W(T)=\identity_N +2[\exp(V^{1/2}/T) -\identity_N]^{-1} ,
\end{equation}
and $\identity_N$ denotes the $N\times N$ identity matrix. In the
ground state case $W(0)$ is given by the identity matrix and so
$\gamma(0)=V^{-1/2}\oplus V^{1/2}$.

An analytical expression for the entanglement (quantified by the
log-negativity $E_l$ \cite{VidWer}) between two complementary groups of
oscillators, $A$ and $B$, was given in terms of the covariance matrix
of the state, which can be written, in turn, only in terms of the
matrix $V$. Then one gets the general formula for the log-negativity
of a thermal state at temperature $T$:
\begin{equation}
E_l=\sum_{k=0}^{N-1} \log_2
\{\max[1,\lambda_k(Q)]\},
\label{logneg}
\end{equation}
where $Q=P\,\omega^-\, P\omega^+$ and
$\omega^\pm=W(T)^{-1}V^{\pm\frac{1}{2}}$. We denoted by
$\{\lambda_k[Q]\}_{k=0}^{N-1}$ the spectrum of the matrix $Q$, whereas
$P$ is an $N\times N$ diagonal matrix with the $i$-th entry given by
$1$ or $-1$ depending on which group, $A$ or $B$, oscillator $i$
belongs to.





\begin{figure} {\includegraphics[width=0.5\textwidth]{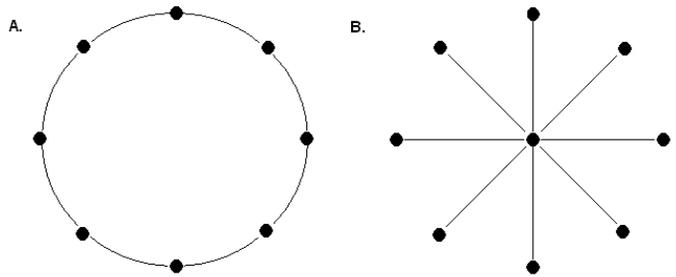}}
 \caption{Pictorial representation of the considered configurations.
   A. Particles interacting via nearest-neighbors couplings. B. Star
   configuration: a central particle interacts equally with the rest
   of the particles.  }
\label{configs}\end{figure}

In Ref. \cite{nos} we considered the thermal states of
Hamiltonian \eqref{Hosc} with a circulant potential matrix $V$ given
by
\begin{equation}\label{Vnn}
V={\rm circ}(1,-c,0,\dots,0,-c)\,.
\end{equation}
\ie, the particles interact via nearest-neighbors interactions (see
Fig. \ref{configs}A). We analyzed the entanglement behavior of the
even-odd and half-half partitions while the system's size $N$ is
increased. A strict area law of type II for the log-negativity was
analytically obtained for large $N$, specifically the entanglement for a given
partition changes proportionally to the area \cala.  The change of the
system temperature just affects the rate the entanglement increases
with $N$ for an even-odd partition (${\cal A}=N$ in this case) and the
entanglement saturation value for the half-half partition (${\cal
  A}=2$). In particular, the threshold temperatures does not depend on
\cala, compatibly with a strict area law of type II.

\begin{figure} {\includegraphics[width=0.4\textwidth]{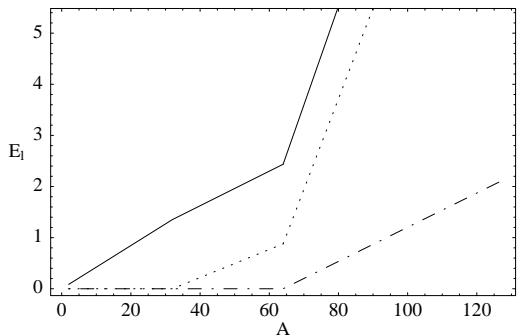}}
 \caption{Log-negativity as a function of the area in the case of
   nearest-neighbor harmonic ring of fixed size $N_{tot}=2^7$ and
   $c=0.4$.  From top to bottom the inverse temperature $\beta=1/T$ is
   given by $\beta=2.5, 2.4, 2$. The system has been partitioned into
   symmetric alternate blocks (see text for details).}
\label{ALaw_a.eps}\end{figure}

We now investigate the area law of type I, that is for fixed
$N=N_{tot}$ and varying the area by changing the geometry of the
partition. Let us focus on two different ways of partitioning the
chain. First, we consider the partition of the system consisting of
$N_{tot}=2^{n}$ particles into $2^{n_b}$ alternate blocks. The area
associated to such a partition is given by ${\cal A}=2^{n_b}$. For
$n_b=n$ one retrieves the even-odd partition and for $n_b=1$ the
half-half partition.  In Fig.  \ref{ALaw_a.eps} we depicted the
log-negativity as a function of the area for $N_{tot}=2^7$ and $c=0.4$
at different temperatures. One sees that, for some fixed temperatures,
there are partitions with PPT while others with non positive partial
transposition, meaning that not all of them become PPT at the
same temperature. A strict area law of type I is then violated,
allowing for the presence of bound entanglement.

\begin{figure} {\includegraphics[width=0.42\textwidth]{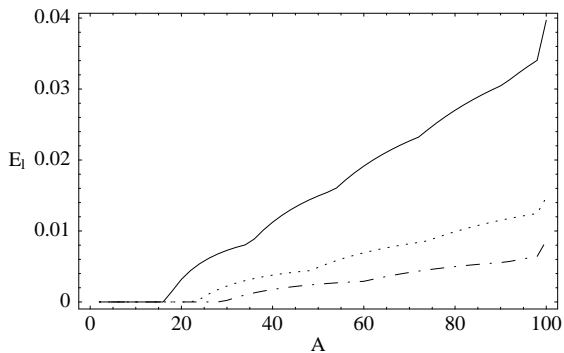}}
 \caption{Same as Fig.~\ref{ALaw_a.eps}, but partitioning the system
   in a non symmetric way (see text for details). The inverse
   temperature $\beta=1/T$ is given by $\beta=1.87, 1.865, 1.863$,
   from top to bottom.}
\label{ALaw_b}\end{figure}

We also considered another way of gradually partitioning the system,
again for fixed system size. Starting from an even-odd partition
(${\cal A}=N_{tot}$) we took one particle at a time, and transferred
it from, say, the even block to the odd block. The process ends with
one of the blocks composed by only one particle (${\cal A}=2$). In
this way the area is decreased by 2 at each step.  In Fig.
\ref{ALaw_b} we depicted the log-negativity as a function of the area
for $N_{tot}=100$, $c=0.4$ and different temperatures. Again one sees
that by increasing the temperature, some partitions become PPT while
others are not.

Through these examples we can see that a strict area law of type I
does not hold. However, notice that even if the entanglement does not
vary strictly proportionally to the area of the subsystem, it still
increases with the area. The details of such a behavior of course
depend on many factors, such as the way the system is partitioned and
the entanglement measure. Nevertheless a general feature seems to be
independent of the partitioning: the more the interaction bonds
intercepted by the partition the more the entanglement across it.

Summarizing, we have seen that in the case of a harmonic nearest-neighbor
ring a strict area law of type I is violated, allowing for the
presence of bound entanglement, whereas a strict area law of type
II is valid, ensuring that bound entanglement survives for large
systems. In the next section, we show that a similar behavior
is also found for spin rings.

\begin{figure}
  {\includegraphics[width=0.5\textwidth,height=0.3\textwidth]{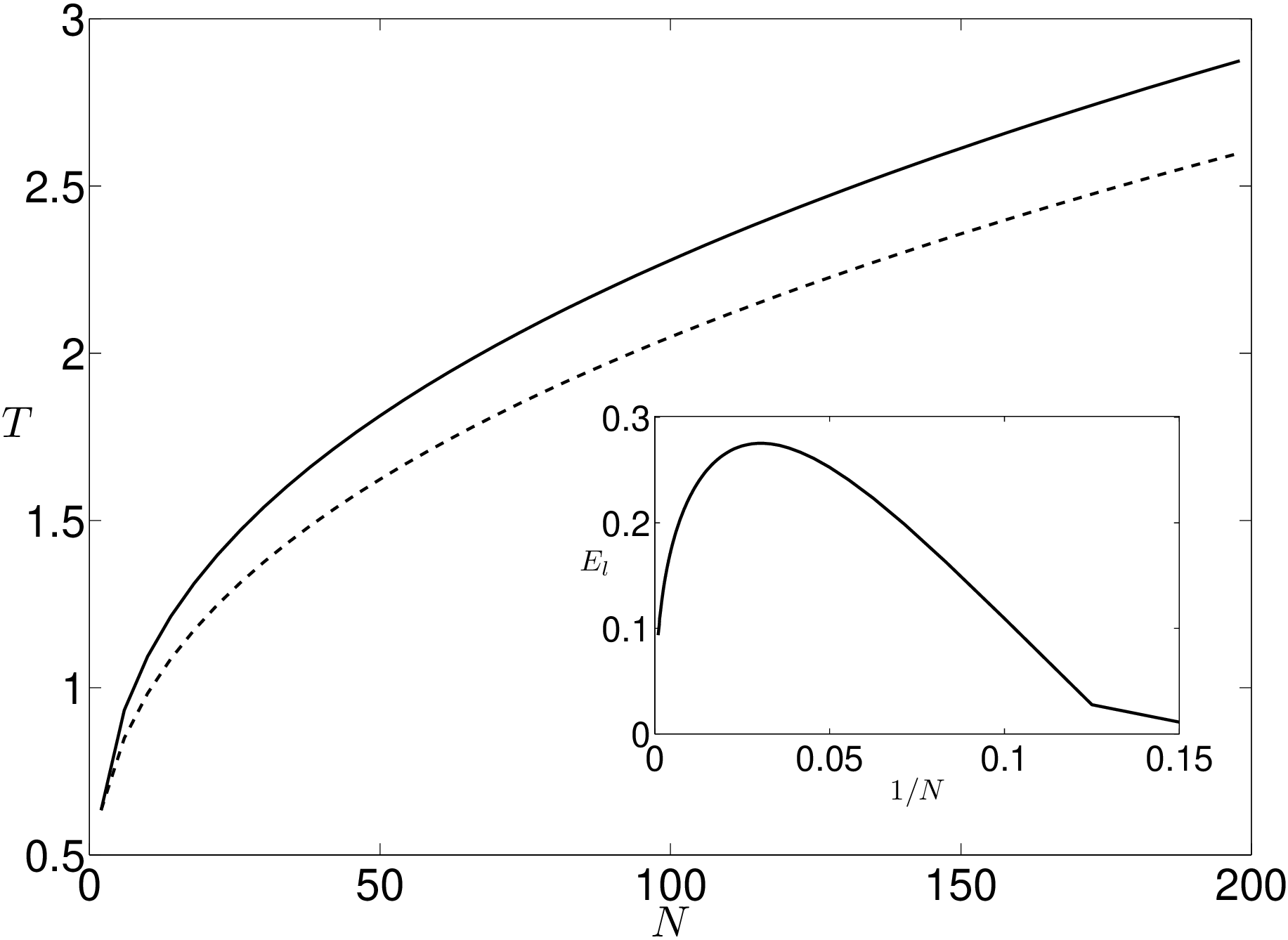}}
  \caption{$T^{c:o}_{th}$ (solid line) and $T^{h:h}_{th}$ (dashed
    line) as a function of the number of oscillators $N$ for the star
    system with interaction $V^{\rm st}$ (see text for details). The
    coupling constant is $c=1$ and similar behaviors have been found
    for different values of $c$. We see that the range of temperatures
    for which the state is bound entangled varies with $N$ for this
    system.  {\bf{Inset:}} Log-negativity of the central vs. rest
    partition as a function of $1/N$, for $c=1$ and $\beta=1$. Notice
    that for large $N$ the log-negativity tends to vanish (see text). The
discontinuity simply comes from the fact that $N$ is integer.}
\label{PhDst}\end{figure}

Let us now turn to a system in which a strict area law of type II is
not valid either. As anticipated, we studied a different topology,
namely a star configuration (see Fig \ref{configs} B). The system is
described by the Hamiltonian \eqref{Hosc} with potential matrix given
by $V^{\rm st}_{11}=1+(N-1)c$, $V^{\rm st}_{ii}=1+c$, $V^{\rm
  st}_{1i}=-c$ and $V^{\rm st}_{ij}=0$ otherwise ($2\le i\le N$,
$c>0$), \ie, all the oscillators are equally connected to a central
one. Clearly, translational symmetry does not hold anymore.
The area law of type I is violated also in this case, allowing to find
a temperature range in which the state is bound entangled. In
Fig.~\ref{PhDst} we depicted how the threshold temperatures for which
the log-negativity is zero, $T^{h:h}_{th}$ (half-half partition) and
$T^{c:o}_{th}$ (central particle versus the outer ones), vary with $N$.
In the region between these two curves, bound entangled states are
present. Nonetheless this is not enough to guarantee that bound
entanglement survives for a large number of particles. First, notice
that the range of temperature $T^{h:h}_{th}-T^{c:o}_{th}$ is no longer
constant with $N$, namely, a strict area law of type II is not valid
in this case. Second, our numerical calculations suggest that the
entanglement between the central particle and the rest (which is the
largest for this configuration) goes to zero as $N\rightarrow\infty$.
Actually, an analytical expression for the log-negativity can be
guessed for $T=0$. In this case, being the state pure, the information
about the entanglement is completely given by the reduced covariance
matrix $\gamma_{\rm red}$ of the central particle. The latter is
simply given by the elements of $\gamma(0)=V^{-1/2}\oplus V^{1/2}$
corresponding to the central particle itself. By calculating
explicitly $\gamma_{\rm red}$ for a small number of particles $N$, one
can recognize the following structure:
\begin{equation}\label{gred}
  \gamma_{\rm red}=
\left(
\frac{1}{N}+\frac{N-1}{N}\sqrt{1+Nc}
\right)\oplus
\left(
\frac{1}{N}+\frac{N-1}{N\sqrt{1+Nc}}
\right)\,.
\end{equation}
The negativity $E_N$ between the central particle and the rest is now
simply given by
\begin{equation}\label{negcr}
E_N=\max[0,\frac{1-\nu}{\nu}]\,,
\end{equation}
where $\nu=\sqrt{\Delta}-\sqrt{\Delta-1}$ and $\Delta$ is the
determinant of $\gamma_{\rm red}$ (see, \eg, Ref.~\cite{Napoli}).
Assuming now that the structure given in Eq.~(\ref{gred}) holds for a
generic $N$ we can extrapolate the behavior of the negativity in the
macroscopic limit. In fact, since $\Delta\rightarrow1$ in the limit
$N\rightarrow\infty$, the negativity itself goes to zero. Considering
now the generic case at $T>0$, it is then reasonable to expect that
the log-negativity goes to zero too for a large number of particles.
As said, our numerical calculations confirms this intuition (see also
the inset in Fig.~\ref{PhDst}). As a consequence, the system is fully
PPT in the macroscopic limit. Such a configuration gives then a
non-trivial example for which the absence of a strict area law of type
II does not allow to clearly identify a range of temperatures for
which bound entanglement is present in the macroscopic limit \cite{noteppt}.

\section{Spin systems}

The scope of this section is to extend the previous analysis to spin systems. We concentrate on the thermal state of systems
composed by $N$ spin-$\frac{1}{2}$ particles, interacting with the
Hamiltonian
\begin{equation}
\label{Hspin}
H=-\sum_{<i,j>}
(\sigma_{x}^{i}\sigma_{x}^{j}+\sigma_{y}^{i}\sigma_{y}^{j})+h\sum_{i=1}^{N}\sigma_z^i\,,
\end{equation}
where the pairs of indices $i$ and $j$ over which we sum define the
topology of the system.

\begin{figure} {\includegraphics[width=0.5\textwidth]{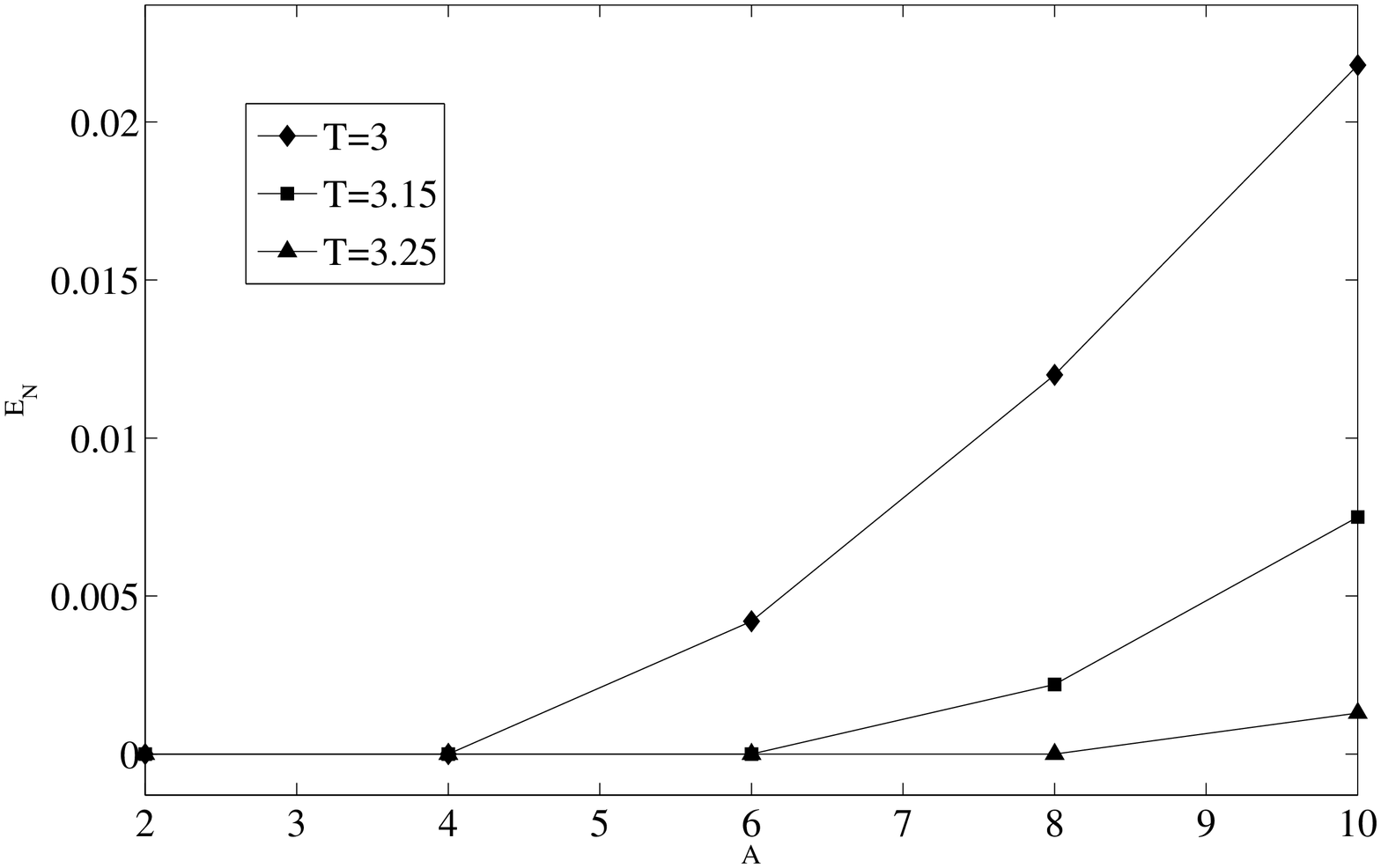}}
  \caption{For a system of $N=10$ spins with nearest-neighbor
    interaction with hamiltonian given by Eq.(\ref{Hspin}) and
    $h=1.9$, we plot the negativity vs. the area of the boundary
    between the two regions, for $T = 3, 3.15, 3.25$. For certain
    values of $T$ we only observe PPT entanglement above a certain
    number of bonds connecting the two regions.}
\label{spinA}
\end{figure}

In the nearest-neighbor configuration (see Fig.\ref{configs}A) we
proceeded as we did for Fig.\ref{ALaw_b} and progressively increase
the boundary area between two regions, starting in the half-half
partition and changing particles from one partition to the other.  In
this situation, for a chain with $N=10$, we can observe
(Fig.\ref{spinA}) again the presence of bound entanglement, since we
have that some partitions are entangled while others are PPT at the same
temperature. This again means that a strict area law of type I
is not valid. On the other hand the validity of an area law of type II
has been numerically shown up to twelve particles in Ref.~\cite{nos},
in analogy to what we have seen for the harmonic ring. This features
strongly support the existence of bound entanglement in the
macroscopic limit also for spin rings.

\begin{figure} {\includegraphics[width=0.4\textwidth]{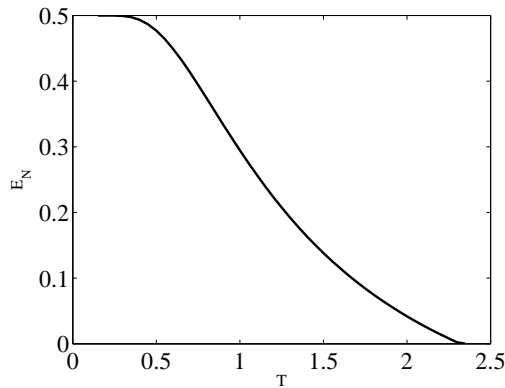}}
  \caption{Negativity for a partition consisting of the central
    particle vs. the others as a function of the temperature $T$ for a
    star system of $N = 4, 6, 8,10$ particles. The Hamiltonian is
    given by Eq.(\ref{Hspin}) with $h=0$. We can see that the
    negativity for different $N$'s coincides, showing its independence
    on the system size.}
\label{spinM}\end{figure}

\begin{figure} {\includegraphics[width=0.5\textwidth]{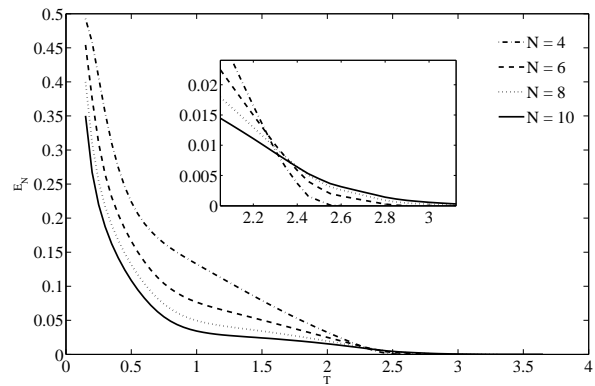}}
  \caption{The negativity for a partition containing only one of the
    external particles vs. the others as a function of the temperature
    $T$ for a star system of $N = 4, 6, 8,10$ particles, where the
    interactions are described by Eq.(\ref{Hspin}) with $h=0$. The
    inset shows in detail the crossing section, and the different
    temperatures where the partition becomes PPT.}
\label{spinE}\end{figure}

Let us now move to a system with a star configuration as in
Fig.\ref{configs}B, and consider the negativity corresponding to
partially transpose either the middle particle or one of the outer
particles. A remarkable feature of this system is that the ground
state entanglement for both partitions is the same for any fixed
$N$, actually these partitions are both maximally entangled. This
can be easily seen by recalling the explicit expression of the
ground state given in Ref.~\cite{bose}. One may then wonder
whether such a behavior holds also at non-zero temperature, which
would suggest that a strict area law of type I is valid. Our
calculations show that this is not the case. Again, different
partitions become PPT at different temperatures, as can be seen
in Figs. \ref{spinM} and \ref{spinE}. Notice that the
central particle now becomes PPT at lower temperatures with
respect to the external ones. Nevertheless, one can conclude the
presence of bound entanglement also in this case by computing the
threshold temperature for a half-half partition.

Another interesting feature of this system is that, by symmetry reasons,
the entanglement between the middle particle and the external
spins is independent of the system size (see Fig.\ref{spinM}). This is
due to the fact that all the eigenstates of this system are of the form \cite{bose}:
\begin{equation}
\frac{1}{\sqrt{2}}(\ket{0}\ket{\alpha_{m,j}}\pm\ket{1}\ket{\alpha'_{m,j}})\,
\end{equation}
where the first ket correspond to the central particle, whereas the
second one to the external particles. The states $\ket{\alpha_{m,j}}$
and $\ket{\alpha'_{m,j}}$ are orthonormal eigenvectors of a high
dimensional fictitious spin. The key point here is that the partial
transpositions with respect to the central particle do not change the
structure of these eigenvectors and this is true for any $N$. As a
consequence, once expanded the thermal state in the eigenbasis above,
one can see that $T^{c:o}_{th}$ does not depend on $N$. In other words
a strict area law of type II holds for this partition.  Notice however
that this behavior does not hold in the case of other partitions, as
we can see in Fig.\ref{spinE} for the case of one external particle
with respect to the rest. In particular, the threshold temperature for
this partition increases with the system size. This fact, considering
that the threshold temperature is size-independent for the middle
particle partition, suggests that the temperature gap for which bound
entanglement appears between the two partitions increases with the
system size.  Recall that this gap appeared constant between the
half-half and even-odd partitions \cite{nos} in systems with
nearest-neighbor interaction.

We can also check the negativity corresponding to transpose one of
the external particles, for different systems sizes at a fixed temperature. In this
scenario, one can see a peculiar behavior: for low temperatures the entanglement for the
central vs. the rest partition can decrease as the system
increases, but for higher temperature the opposite holds.
The crossing temperature appears clearly
non-trivial, between 2.2 and 2.4, see the inset of Fig.\ref{spinE}. Again
the entanglement in different partitions do not vanish at the
same temperature, allowing the presence of bound entanglement.

Before ending this section, let us stress that our numerical
calculations are restricted to small number of particles due to
computational hardness. Although not shown here, similar results can
be found for other types of interactions, e.g. using Heisenberg-type
hamiltonians.

\section{Conclusions}

To conclude, we have extended the results of Ref. \cite{nos} and
considered different ways of studying the entanglement distillability
properties of thermal states of many-body systems. We have considered
systems of harmonic oscillators and spin-one-half particles in a chain
and star topology. In general, our results show that a strict
entanglement area law of type I (when changing the partitions for a
fixed system size) is not fulfilled.  Since the different partitions
become PPT at different temperatures, bound entanglement appears for a
temperature range in a natural way. Concerning the preservation of
this range of temperatures when the system size is increased, we
pointed out that a different approach to area laws should be
addressed. In particular, an entanglement area law of type II (when
changing the system size) is then useful to prove the presence of
bound entanglement in the macroscopic limit of an infinite number of
particles.



\begin{acknowledgements}

This work is supported by the EU QAP project, the Spanish MEC, under
  FIS2004-05639 and Consolider-Ingenio QOIT projects, and a ``Juan de
  la Cierva" grant, the Generalitat de Catalunya, and the Universit\'a
  di Milano under grant ``Borse di perfezionamento all'estero".
\end{acknowledgements}

\end{document}